\documentclass[twocolumn,showpacs,preprintnumbers,amsmath,amssymb]{revtex4}

\usepackage{graphicx}
\usepackage{dcolumn}
\usepackage{bm}

\begin{document}


\title{Controlling the efficiency of spin injection into graphene by carrier drift}

\author{C. J\'{o}zsa$^1$}
\author{M. Popinciuc$^2$, N. Tombros$^1$, H.T. Jonkman$^2$}
\author{B.J. van Wees$^1$}

\affiliation{$^1$Physics of Nanodevices, Zernike Institute for Advanced Materials, University
of Groningen, The Netherlands\\
$^2$Molecular Electronics, Zernike Institute for Advanced Materials, University
of Groningen, The Netherlands
}

\date{\today}

\begin{abstract}

Electrical spin injection from ferromagnetic metals into graphene is hindered by the impedance mismatch between the two materials. This problem can be reduced by the introduction of a thin tunnel barrier at the interface. We present room temperature non-local spin valve measurements in cobalt/aluminum-oxide/graphene structures with an injection efficiency as high as 25\%, where electrical contact is achieved through relatively transparent pinholes in the oxide. This value is further enhanced to 43\% by applying a DC current bias on the injector electrodes, that causes carrier drift away from the contact. A reverse bias reduces the AC spin valve signal to zero or negative values. We introduce a model that quantitatively predicts the behavior of the spin accumulation in the graphene under such circumstances, showing a good agreement with our measurements.

\end{abstract}

\maketitle

  The predictions of a long spin relaxation time in graphene~\cite{TheorSpinGraphene}, together with the availability of micrometer sized graphene flakes~\cite{RiseofG} fueled a number of experimental studies on graphene spin valve type devices. Since 2006, several successful spin injection experiments were reported in field effect transistor geometries of lateral dimensions of a few micrometers~\cite{Tombros2007,Hill2006,Cho2007,Nishioka2007,Ohishi2007,Kawakami2008,Goto2008}, all relying on electrical spin injection and detection using ferromagnetic metal electrodes. In earlier experiments~\cite{Tombros2007} we have determined the spin polarization of our ferromagnet/aluminum-oxide/graphene contacts as 10\% (room temperature experiments), strongly depending on the contact resistances i.e. the properties of the tunnel barrier. We have also presented a method to manipulate the spin transport through graphene from injector to detector by means of carrier drift, under the influence of a DC electric field~\cite{JozsaDrift}.

At this point the question arises, what determines the efficiency of spin injection from the ferromagnet into graphene. In the limit of clean metal on graphene, experiments show that an ohmic contact is formed at the interface, and spin injection is determined by the spin selective resistivity of the ferromagnet. Due to the impedance mismatch between the two materials~\cite{ImpMis1,ImpMis2}, this leads to a very inefficient injection of the spin polarized current into the graphene. In case of a tunnel barrier, on the other hand, the spin dependent tunneling takes the role of the spin-dependent contact resistance, and impedance mismatch can be reduced~\cite{ImpMis3}. This way we can define an intrinsic spin polarization of a contact $P$ that is determined by the nature of the interface/tunnel barrier and the ferromagnet, and an effective injection (or detection) efficiency $P_{inj/det}$ that takes into account the presence of graphene and is lower due to the impedance mismatch.

\begin{figure}[b!]
\includegraphics[width=8cm]{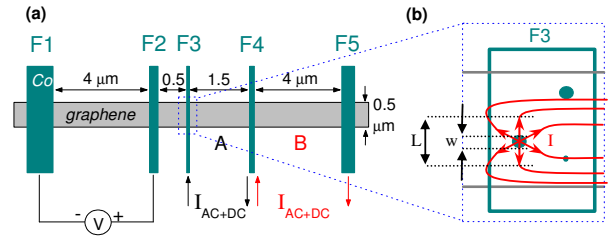}
\caption{\label{fig:device} (color online) (a) The sample layout and non-local measurement geometry, injection schemes A (black) and B (red). The direction of arrows indicate positive bias. (b) Illustration of the carrier drift on a length scale L from a pinhole of width w. Under DC bias, a strong local electric field induces drift of carriers in the graphene (see arrows) facilitating or blocking the spin injection from the ferromagnetic electrode F3 through the pinhole.}
\end{figure}

A study of the interplay of mechanisms behind electrical spin detection in lateral Fe/GaAs structures (Schottky tunnel barriers) was done recently by Crooker et al.\cite{Crooker2008}. The authors found that the sensitivity of such a spin detection scheme can be tuned by an electrical bias on the interface. The effect was explained by the bias dependence of the tunneling spin polarization (interface effect) as well as the bias dependence of the spin transport in the GaAs before the spins reach the detection point (bulk effect).

Here we present the manipulation of the effective spin injection efficiency at electrically biased ferromagnet/tunnel barrier/graphene interfaces, where the tunnel barrier is below 1 nm thick with (probably) relatively transparent pinholes. The four-terminal spin valve devices built for this purpose, illustrated in Fig.~\ref{fig:device}(a), are similar to the ones investigated in Ref.~\onlinecite{Tombros2007} and~\onlinecite{JozsaDrift}; see these references for a detailed description of the fabrication procedure. In the non-local measurement geometry we employ the current injection circuit (F3, F4, F5) is separated from the voltage probes (F1, F2) decoupling the charge current from the spin current, a technique that allows detection of a voltage difference that can only be attributed to a spin diffusion through the graphene layer\cite{Zutic}. The spin injector/detector electrodes are 50 nm thick Co strips of widths (left to right) 800, 250, 90, 140 and 350 nm. The different magnetic shape anisotropies yield switching fields from below 20 mT (F1, widest) up to 85 mT (F3, narrowest). The Co contacts patterned by electron beam lithography are separated from the graphene by a 0.8 nm thick Al$_2$O$_3$ layer; the distances between them are shown on the figure.  The contact resistances probed by three-point electrical measurements were between 50 k$\Omega$ and 90 k$\Omega$. The fact that the contact resistances do not scale with the contact areas, as well as the granular morphology of the sub-nanometer aluminum-oxide layer indicates an electrical contact mediated by a number of pinholes. The spin injection and detection method was based on a standard low-frequency AC lock-in technique with a current of 0.5~$\mu$A RMS. In addition to the AC injection current, a DC current bias (-5 to 5 $\mu$A) was applied on the current injector electrodes F3, F4 and F5, in a geometry explained below.

In the first set of experiments we used contacts F3 and F4 as AC+DC current injectors, and contacts F2 and F1 as the non-local voltage probe (scheme A on Fig.~\ref{fig:device}(a)). As a consistency check, we performed a second set of experiments on the sample, where everything was kept the same except the current injectors, that we shifted to the electrodes F4 and F5, leaving electrode F3 floating (scheme B). This resulted in an opposite DC bias on the electrode F4 in comparison with scheme A. Furthermore, a second sample was manufactured with the same spin valve geometry but contact resistances in the 1 - 10 k$\Omega$ range, where we have performed similar measurements. The results support the data we present in this manuscript, indicating good reproducibility of the effect.

\begin{figure}[b!]
\includegraphics[width=8.5cm]{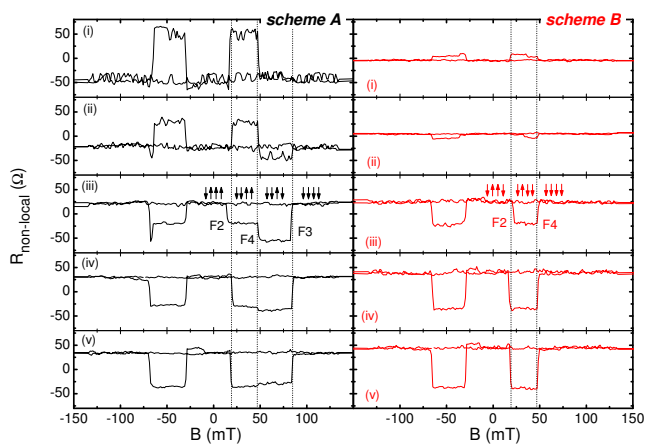}
\caption{\label{fig:meas} (color online) Non-local spin valve measurements at DC current biases -5, -2.5 , 0, 2.5 and 5 $\mu$A [panels (i) to (v)], for contact scheme A (left column) and B (right column). In scheme A, the magnetic switching of contacts F2, F3 and F4 is visible (also shown by black arrows).}
\end{figure}

In these experiments we swept the magnetic field aligned parallel to the electrodes from -150 to +150 mT, while monitoring the non-local resistance defined as $R_{\mbox{nl}} = V_{AC}/I_{AC}$. This was repeated for a number of selected bias currents $I_{DC}$. The two measurement sets for schemes A and B are shown on Fig.~\ref{fig:meas}. Every step in the spin valve measurements, as indicated for the case of the zero bias curves (iii), corresponds to the magnetization switch of an electrode. Note, that electrodes F1 and F5 were at a distance approximately 3 times greater than the spin diffusion length in graphene ($\lambda_{sf} \simeq 2\mu$m \cite{Tombros2007}), therefore their contribution to the measured resistance was too weak to appear in the plotted data. By electrostatic measurements the graphene was determined to be p-type, carrier density being in the $10^{16}$ m$^{-2}$ range (at zero applied gate voltage).

Examining the zero-bias measurement in scheme A (panel iii left), we can follow the evolution of the resistance while sweeping the magnetic field from zero to +150 mT. We start with all four electrodes F1-4 aligned parallel ("up" direction) in their magnetization. In this scheme, electrode F3 injects spin-up polarized current and creates a spin imbalance in the graphene. Electrode F4 extracts spin-up, i.e. creates an opposite spin imbalance. The spin imbalance diffuses through the graphene and arrives to the detector F2 (sensitive to the spin-up channel). Since F3 is much closer to the detector than F4, its effect on the detector will be stronger and we measure a positive non-local resistance of approximately 25$\Omega$. When the magnetic field reaches the value of 20 mT, we see a step in the resistance that we associate with the switching of detector F2 to the "down" direction, being now antiparallel with the injector F3. F2 probes thus the "down" spin channel, yielding a negative resistance level. At approximately 50 mT, the injector electrode F4 switches and another step appears in the resistance. The two injectors F3 and F4 are now antiparallel oriented, which means they both inject spin up carriers. The switching of F4 therefore further increases the signal on the detector electrodes, this becoming more negative. Finally, at 85 mT field the injector F3 switches to the "down" state. All four electrodes being parallel again, the measured resistance shows the initial value of +25$\Omega$.

On the right panel (iii) of Fig.~\ref{fig:meas}, in the injection scheme B, we see only two steps, at magnetic fields of 20 and 50 mT. Electrode F3 being not connected, it does not contribute to the signal. The four vertical arrows represent the magnetic orientation of F1, F2, F4 and F5. The steps associated with the switching of electrodes F2 and F4 happen at the same magnetic fields both in scheme A and B, indicating a consistent behavior of the spin valve.

\begin{figure}[t!]
\includegraphics[width=8.5cm]{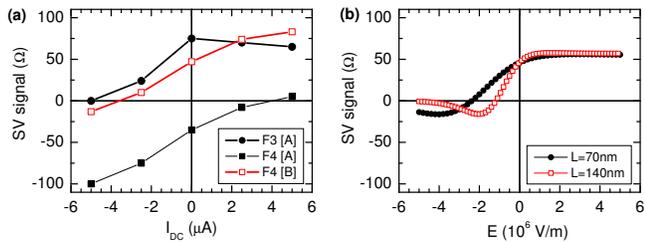}
\caption{\label{fig:measimu} (color online) (a)The resistance steps associated with the switching of electrodes F4,F3 in the schemes A and B, plotted as a function of the applied DC current bias; (b) simulated with the drift model for two different drift lengths $L$.}
\end{figure}

The non-local resistance levels we measure depend on the relative magnetic orientation of the electrodes controlled by the magnetic field, on the distance between injectors and detectors fixed by design, and on their spin injection/detection efficiencies. The efficiencies are correlated to the spin polarization of the contacts between graphene and injectors $P_{inj}$ and, respectively graphene and detectors  $P_{det}$. These can be calculated from the spin valve measurement using the relation from Ref.~\onlinecite{Tombros2007}
\begin{equation}
R_{\mbox{nl}} = \frac{P_{inj} P_{det}  \lambda_{sf}}{2W_g \sigma} \mbox{exp}(-L_{sv}/\lambda_{sf}),
\end{equation}
where the non-local resistance $R_{\mbox{nl}}$ is given by the measurements, $\lambda_{sf} \simeq 2 \mu$m is the spin diffusion length in graphene~\cite{Tombros2007}, $W_g = 500$nm is the width of the graphene channel, $\sigma = R_g^{-1} \simeq 1.1 \cdot 10^{-3} \Omega^{-1}$ is its measured conductivity and $L_{sv}$ is the injector-detector distance (F3 to F2 and F4 to F2, respectively). For our case, the spin polarization this formula yields is 25\%.

Applying an electric bias on the injector electrodes F3 and F4 (scheme A) and respectively, F4 and F5 (scheme B) changes the picture. On Fig.~\ref{fig:meas} panels (i), (ii), (iv) and (v) we plotted the spin valve measurements taken at DC bias currents of -5, -2.5, +2.5 and +5 $\mu$A. Through the measurement series (i) -- (v) in scheme A we can follow the behavior of the resistance steps (thus the spin injection/detection efficiencies) associated with the electrodes F2, F3 and F4 under the applied biases. In case of injector electrode F3, a positive bias (panels iv, v) does not seem to change much in the stepsize; it stays +75 $\Omega$.  However, when the bias is reversed, the resistance step is gradually reduced to zero. This indicates, that under negative bias, the spin injection of this contact becomes very inefficient. The behavior of the injector electrode F4 is consistent with its opposite bias compared to F3. The resistance step associated with it is increasing from -100 $\Omega$ to +10 $\Omega$ when we change the bias from -5 $\mu$A to +5 $\mu$A. Note, that the steps associated with the efficiency of F4 are generally lower, due to the larger injector-detector distance, but they follow approximately the same trend as in the case of F3. The sensitivity of detector F2 is unchanged during the measurements (no bias applied on the detectors); the resistance steps associated with its switching are simply equal to the difference of the two steps caused by F3 and F4. For a better overview, on Fig.~\ref{fig:measimu} (a) we have summarized these observations, plotting the behavior of the resistance steps against the current bias. The measurements done in scheme B deliver an additional curve for F4; this electrode is now under positive bias, and the sign of both the spin valve signal as well as $I_{DC}$ is reversed.

 These experiments indicate, that biasing the injector electrodes yields a dramatic change in their spin injection efficiencies, enhancing the spin valve signal to a saturation value or suppressing it completely. Applying Equation (1) for the spin valve signal we measured in scheme B, at +5 $\mu$A bias we calculate a spin polarization $P_{inj} = 43$\% for the contact between F4 and graphene. To do this, we keep the spin polarization of the unbiased detector/graphene contact  $P_{det}$ at the original 25\%. In case of maximum reverse bias on the other hand, the AC measurements show a reversed spin valve behavior (see panels (i) on Fig.~\ref{fig:meas}) that will be addressed later.

The physics behind spin injection through an inhomogeneous tunnel barrier under the action of a DC electric field can be modeled and understood considering a strong local drift of charge carriers. This happens in the graphene region directly around the injection points (the pinholes) on a characteristic length scale $L$. The width of the pinholes $w$ in the Al$_2$O$_3$ barrier is another important parameter that defines the electric field $E$ possible to be generated by a current $I_{DC}$ sent through the contact. An illustration of the idea is included in Fig.~\ref{fig:device} (b).

A qualitative picture of the effect can be created in the following way. From injection to detection, there are three electrically coupled regions of different spin transport: injection from the ferromagnet through the tunnel barrier via pinholes, spin transport through the graphene in the immediate vicinity of the injection point (drift region) and finally, diffusion/relaxation towards the detection point. In the drift region the DC bias gives rise to the electric field $E$. This yields a drift-diffusion type of transport as described in Ref.~\cite{JozsaDrift} , however in this case on the short length scale $L$. The carriers in the graphene drifting away from the injection point (in case of positive DC bias and p-type graphene) reduce the backflow of spins and thus facilitate further injection of spin polarized current through the pinhole(s). The upper limit of the effect is a measurement of the intrinsic spin polarization $P$ possible to inject from the ferromagnetic electrodes (large $E$), when impedance mismatch is eliminated. On the other hand, an opposite electric field polarity (or carrier type) will result in a carrier drift towards the injection point and therefore, it will keep the local spin polarization high, reducing the efficiency of the spin-polarized injection.  An increased negative bias enhances the impedance mismatch to a point when the drift starts to dominate the spin transport. In our AC measurements, this yields a negative differential resistance (lower spin injection for stronger electric fields) and thus the reversed AC spin valve signals on Fig.~\ref{fig:meas}. Finally, above a threshold electric field value the strong drift effect prevents detection of any AC spin signal at the electrodes F2,F1.

To quantitatively model the transport in the drift region, we consider a one-dimensional drift-diffusion of the spins along the $x$-axis. We write the spin accumulation $n_s$ along the axis $x$ ($x=0$ represents the injection point) as described in Refs.~\cite{YuFlatte1,JozsaDrift} in the form of
\begin{equation}
n_s(x) = A \exp{(+\frac{x}{\lambda_+})} + B\exp{(-\frac{x}{\lambda_-})},
\end{equation}
where $\lambda_{\pm}$ are the up/downstream spin transport length scales as in~\cite{YuFlatte1}. The spin current density flowing through the region due to diffusion and drift can be written as
\begin{equation}
j_s(x) = -D\frac{dn_s(x)}{dx} + v_D \cdot n_s(x).
\end{equation}
where $v_D$ represents the drift velocity.

The coupling at the edges of the drift region yields two boundary conditions. Injection from the Co electrode is modeled by a spin-polarized current source $I_s = P \cdot I$ -- where $P$ represents the intrinsic spin polarization without any impedance mismatch -- parallel with a spin flip resistor $R_s$ of dimension $m^{-1} \cdot s$. This contains the contact resistance $R_c \simeq 50$ k$\Omega$ and the impedance mismatch between the graphene and the ferromagnet: $R_s = 4R_c W_g N_{2D} e^2$, where $W_g = 500 $nm is the graphene width and $N_{2D}$ is the 2-dimensional density of states in graphene. The relation $N_{2D}e^2 = D/R_g$ allows us to determine $R_s$ from the measured values of the graphene sheet resistance $R_g \simeq 875\Omega$ and diffusion constant $D \simeq 0.02 $m$^2$/s. The spin current density entering  the drift region is therefore the spin current of the source $I_s$ minus the relaxation through the contact represented by $R_s$.

The spin current density exiting the drift region is coupled to the other end of the device, where spin transport is governed by diffusion and relaxation, and we model it with another spin flip resistor $R_{out} = \lambda_{sf}/2D$. Here the factor 2 is introduced since spin relaxation can take place on both sides of the contact. These considerations allow us to calculate the spin accumulation at the detector electrode, as a function of the electric field $E$ and the drift length $L$. Knowing the spin polarization of the detector electrode $P$, we obtain the AC voltage difference $V_{AC}$ on it that is due to the spin transport. On Fig.~\ref{fig:measimu} (b) we plotted this voltage difference in form of a non-local resistance $V_{AC}/I_{AC}$ against the electric field for two different drift lengths, using the experimental parameters of the injection scheme B. A linear correspondence between the x-axis of Fig.~\ref{fig:measimu}(a, experiment) and (b, simulation) is given by the formula $E = I_{dc} \cdot R_L / L = I_{dc} \cdot R_g / w$, where $R_L$ is the total resistance of the graphene along the drift region $L$ and $w$ is the pinhole size. In this case, we refer to the collective effect of all pinholes in the tunnel barrier that contribute to the injection process. Comparing the simulation to the measurements and considering the length of the modeled drift region $L$ to be around 100 nm, we note that the electric fields necessary to obtain good accordance are in the $\pm 10^6$ V/m range. This results in a drift velocity $v_D = \mu \cdot E \simeq 0.25 \cdot 10^6$ m/s that equals approximately the quarter of the Fermi velocity. This is only possible to achieve with $w<5$ nm. Considering, that we probably have more than one pinhole, the individual average size of them is at the nanometer range or below. This is consistent with the fact, that spin transport under the Co contacts is possible (the graphene is not shorted by large contact areas).

In the presented measurements, the entire graphene channel was p-type. We have done similar sets of measurements where the graphene between the electrodes was n-type and, respectively in the vicinity of the Dirac neutrality point, using electrostatic gating. Aside of a scaling of the signal with the carrier density, the results were consistently the same in all three regimes. This means that gating has no influence on the spin injection process, only on the diffusive transport in the graphene channel between the contacts~\footnote{The gate voltage applied on the graphene was efficiently screened by the proximity of the metal electrode therefore under the ferromagnetic contacts the carrier type was insensitive to gating, remaining p-type. Furthermore, the electrostatic effect of the contact voltage bias on the density underneath the contacts proved not to be enough to change the graphene to n-type.}.

In conclusion, we demonstrated electrical spin injection into graphene at room temperature with a high efficiency of 25\%, controllable by electrical bias on the contacts. Non-local signals up to 100$\Omega$ were obtained this way. The results were explained by injection through nanometer sized pinholes in the tunnel barrier between the metal and graphene followed by strong local carrier drift. We do not fully understand yet the nature of our Al$_2$O$_3$ barrier. The contact resistances we measured indicate the presence of relatively transparent (though not metallic) pinholes; the morphology of the oxide layer has yet to be verified by surface characterization experiments.

This work was financed by the Zernike Institute for Advanced Materials, NanoNed, NWO and FOM.

\end{document}